\documentclass[10pt, two column, twoside]{IEEEtran}
\usepackage{color}
\usepackage[colorlinks, linkcolor=color1, anchorcolor=blue, citecolor=color1]{hyperref}

\usepackage{amssymb}
\usepackage{amsmath}
\usepackage[lined,boxed,commentsnumbered, ruled]{algorithm2e}
\usepackage{mathrsfs}
\usepackage{algorithmic}
\usepackage{bm}
\usepackage{hhline}

\usepackage{pgfplots}
\usepackage{tikz}
\usetikzlibrary{arrows}
\usepackage{caption}
\usepackage{subcaption}
\usepackage{makecell}
\usepackage{graphicx,booktabs,multirow}

\definecolor{colorhkust}{RGB}{20,43,140}
\definecolor{colortsinghua}{RGB}{116,52,129}
\definecolor{color1}{RGB}{128,0,0}

\date{}

\begin{document}

\title{Mobile Edge Intelligence and Computing for the Internet of Vehicles}
\author{Jun Zhang, Khaled B. Letaief

        \thanks{Jun Zhang is with the department of Electronic and Information Engineering, Hong Kong Polytechnic University, Hong Kong (email: jun-eie.zhang@polyu.edu.hk). Khaled B. Letaief is with the department of Electronic and Computer Engineering, Hong Kong University of Science and Technology, Hong Kong (email:eekhaled@ust.hk). }
                }

\maketitle


\maketitle

\begin{abstract}
The Internet of Vehicles (IoV) is an emerging paradigm, driven by recent advancements in vehicular communications and networking. Advances in research can now provide reliable communication links between vehicles, via vehicle-to-vehicle communications, and between vehicles and roadside infrastructures, via vehicle-to-infrastructure communications. Meanwhile, the capability and intelligence of vehicles are being rapidly enhanced, and this will have the potential of supporting a plethora of new exciting applications, which will integrate fully autonomous vehicles, the Internet of Things (IoT), and the environment. These trends will bring about an era of intelligent IoV, which will heavily depend upon communications, computing, and data analytics technologies. To store and process the massive amount of data generated by intelligent IoV, onboard processing and Cloud computing will not be sufficient, due to resource/power constraints and communication overhead/latency, respectively. By deploying storage and computing resources at the wireless network edge, e.g., radio access points, the \emph{edge information system} (EIS), including edge caching, edge computing, and edge AI, will play a key role in the future intelligent IoV. Such system will provide not only low-latency content delivery and computation services, but also localized data acquisition, aggregation and processing. This article surveys the latest development in EIS for intelligent IoV. Key design issues, methodologies and hardware platforms are introduced. In particular, typical use cases for intelligent vehicles are illustrated, including edge-assisted perception, mapping, and localization. In addition, various open research problems are identified.
\end{abstract}


\section{Introduction}

The automobile industry has been one major economic sector for over a century, and its economical and societal impacts continue to expand. For example, automakers and their suppliers are responsible for 3\% of the US GDP, and no other manufacturing sector generates as many jobs in the US \cite{US}. Due to the increasing convenience, comfort, low cost, and fuel efficiency, there have been more and more vehicles on the road. In 2018, China and the US, the two largest automobile markets, sold 23.2 and 17.2 million passenger cars, respectively \cite{VDA}. The increasing number of vehicles have caused various issues such as traffic congestion, accidents, and air pollution.  According to the \emph{Global status report on road safety 2018} of the World Health Organization (WHO) \cite{WHO}, the number of road traffic deaths has reached 1.35 million in 2016. In other words, there are nearly 3,700 fatalities on the world's roads every day.

Significant efforts have recently been spent on improving vehicle safety and efficiency. In particular, information and communication technologies have been regarded as promising tools to revolutionize vehicular networks. Connecting vehicles via Vehicular Ad-hoc Networks (VANET) represents an early attempt to support safety-related applications, such as accident warning, crash notification, and cooperative cruise control \cite{LuChe14,SieErb18}.  With VANET, neighboring vehicles are allowed to communicate with each other via Vehicle-to-Vehicle (V2V) communications, which helps to improve driving safety. Vehicles can also communicate with the roadside infrastructure via Vehicle-to-Infrastructure (V2I) communications to collect road and traffic-related information. These communication links are enabled by Dedicated Short Range Communications (DSRC) or cellular-enabled Vehicle-to-Everything (V2X) communications \cite{AbbZhu16,ZhaJia18,LiaPen17}. 

The Internet access is not fully available in VANET, which limits the scope of its applications. To extend the capabilities of VANET, the Internet of Vehicles (IoV) has been proposed to form a global network of vehicles, evoking collaborations between heterogeneous communication systems to provide reliable Internet services \cite{LeeGer16,XuZho18}. IoV will have communications, processing, storage, and learning capabilities. In particular, with IoV, vehicles will be able to leverage resources such as Cloud storage and computing. Besides vehicle driving and safety, IoV will also facilitate urban traffic management, vehicle insurance, road infrastructure construction and repair, logistics and transportation, etc. As a special case of the Internet of Things (IoT), IoV shall  be integrated with other systems, such as Smart City.

Meanwhile, we are witnessing the growing intelligence of vehicles, thanks to recent advancements in embedded systems, navigation, sensors, data acquisition and dissemination, and big data analytics \cite{LevJak11,LueHim12}. It started with assisted-driving technologies, namely, Advanced Driver-Assistance Systems (ADAS), including emergency braking, backup cameras, adaptive cruise control, and self-parking systems \cite{BenDie14}. Around the world, the number of ADAS systems rose from 90 million units in 2014 to about 140 million in 2016, a 50\% increase in just two years \cite{McKinsey17}. According to the definitions of autonomous vehicles laid out by the Society of Automotive Engineers (SAE) International, the above systems mainly belong to Level 1 and Level 2 of automation. Tesla's Autopilot system also falls in this category \cite{Tesla}. Automotive manufacturers and technical companies, such as Google, Uber, Tesla, Mobileye, are investing heavily on higher levels of driving autonomation. The 2018 Audi A8 is the first Level 3 self-driving car available in production \cite{Gau}. Predictions vary, but many forecast that autonomous vehicles with Level 4 and Level 5 will be available in the market within a decade.


The upcoming intelligent IoV will need support from various sectors, including automobile, transportation, wireless communications, networking, security, robotics, as well as regulators and policy makers. This survey shall provide a perspective from information and communication technologies on intelligent IoV. In particular, we will advocate that the integration of storage, communications, computing, and data analytics at the wireless network edge, e.g., radio access points, provides an effective framework in addressing the data acquisition, aggregation, and processing challenges for intelligent IoV. In the following, we first elaborate on the big data challenges in intelligent IoV, and motivate the need for an edge information system.

\subsection{Big Data in Intelligent IoV}

The advancements in information technologies, including communication, sensing, data processing, and control, are transforming the transportation system from conventional technology-driven systems into more powerful data-driven intelligent transportation systems \cite{ZhaWa11}. This trendy movement will generate a tremendous amount of data. Over the past two decades, the wireless industry has been struggling with the mobile data explosion brought by smartphones \cite{AndBuz14}. Such struggle will be dwarfed by the expected huge amount of data to be generated by intelligent IoV. Intelligent vehicles are equipped with multiple cameras and sensors, including Radar, Light Detection And Ranging (LiDAR) sensors, Sonar, Global Navigation Satellite Systems (GNSS), etc. It is predicted that there will be more than 200 sensors in future vehicles \cite{ASEE}, with total sensor bandwidth reaching 3 Gbit/s ($\sim$1.4 TB/h) to 40 Gbit/s ($\sim$19 TB/h) \cite{Hei17}. As estimated by Intel, each autonomous vehicle will be generating approximately 4,000 GB of data a day, equivalent of the mobile data generated by almost 3,000 people. Assuming mere 1 million autonomous cars worldwide, then automated driving will be equivalent to the data of 3 billion people. Due to this huge surge, Brian Krzanich, CEO of Intel, remarked that ``data is the new oil in the future of automated driving'' \cite{Krz16}.


The big data generated by intelligent IoV will place unprecedented pressure on communication, storage and computing infrastructures. While onboard computing and storage capabilities are increasing rapidly, they are still limited compared with the scale of data to be stored and processed. For example, NVIDIA's self-driving learning data collection system adopts Solid-State Drive (SSD) as the external storage, up to several Terabyte, which will be filled within hours by sensing data. Furthermore, the computation needed to process these data will easily exhaust the onboard computing resources. A car equipped with 10 high-resolution cameras can generate 2 gigapixels per second of data. Processing that amount of data through multiple deep neural networks converts to approximately 250 TOPS (trillion operations per second) \cite{NVIDIA18}. Meanwhile, to achieve better safety than the best human driver, who takes action within 100$\sim$150 ms, autonomous driving systems should be able to process real-time traffic conditions within a latency of 100 ms \cite{LinZha18}, which demands significant amount of computing power. While power-hungry accelerators such as Graphics Processing Units (GPUs) can provide low-latency computation, their high power consumption, further magnified by the cooling load to meet the thermal constraints, can significantly degrade the driving range and fuel efficiency of the vehicle \cite{LinZha18}.

There have been many proposals for using Cloud computing \cite{ZhaChe10} to help intelligent vehicles, and some of them have already been implemented, such as Cloud-based software update, or training powerful deep learning models \cite{KehPat15}. Cloud computing platforms will certainly be an important supporter for IoV, but they are not sufficient. While cost and power consumption are the main limiting factors for onboard computation, the long latency and the massive data transmission are the bottlenecks for Cloud based processing \cite{ZhaWan18}. The round trip time from a mobile client to a cloud center may easily be longer than 100 ms \cite{CarKoc19}. Moreover, such latency highly depends on the wireless channel condition, the bandwidth of the network, and traffic congestion, so real-time processing and reliability cannot be ensured. As shown in \cite{CarKoc19}, considering the latency requirement, if the speech recognition task of a driving assistance system is to be offloaded, the server has to be located at the nearby base station, i.e., at the edge of the wireless network. This is aligned with the recent trend to deploy computing resources at the edge of wireless networks \cite{hu2015mobile,ShiCao16,MacBec17,MaoYou17}, to be elaborated below.

\subsection{Living On The Edge}

To overcome the limited capabilities of onboard computing, communication, storage, and energy, while avoiding excessive latency in Cloud computing, deploying resources at the wireless network edge has received significant attention from both academia and industry. Popular content such as video files, which dominates mobile data traffic, is likely to be repeatedly requested by different users, and such requests are predictable. Thus, deploying storage units and caching popular content at the wireless network edge, i.e., \emph{wireless edge caching}, stands out as a promising solution for efficient content delivery \cite{ShaGol13,BasBen14,BasBen15,JiCai16}. Meanwhile, the revival of artificial intelligence (AI) and the emergence of intelligent mobile applications demand platforms that can support computation-intensive and delay-sensitive mobile computing. \emph{Mobile Edge Computing} (MEC) is an emerging technology that has the potential to unite telecommunications with Cloud computing to deliver cloud services directly from the network edge and support delay-critical mobile applications. This is achieved by placing computer servers at the base stations (BSs) or radio access points \cite{hu2015mobile,ShiCao16,MacBec17,MaoYou17,MaoZha16}. Edge caching and computing platforms further enable \emph{edge AI}, which trains and deploys powerful machine learning models at the edge servers and mobile devices, and has been regarded as a key supporting technology for IoT \cite{WanTuo18,ZhoChe19,IBM,Intel}. Edge AI is changing the landscape of the semiconductor industry. In 2018, shipment revenues from edge AI reached \$1.3 billion, and by 2023 this figure is forecast to reach \$23 billion \cite{Deans}. Collectively, these platforms will be referred to as the \emph{Edge Information System} (EIS) in this paper.

\begin{figure*}[t]
\center
\includegraphics[scale = 0.43]{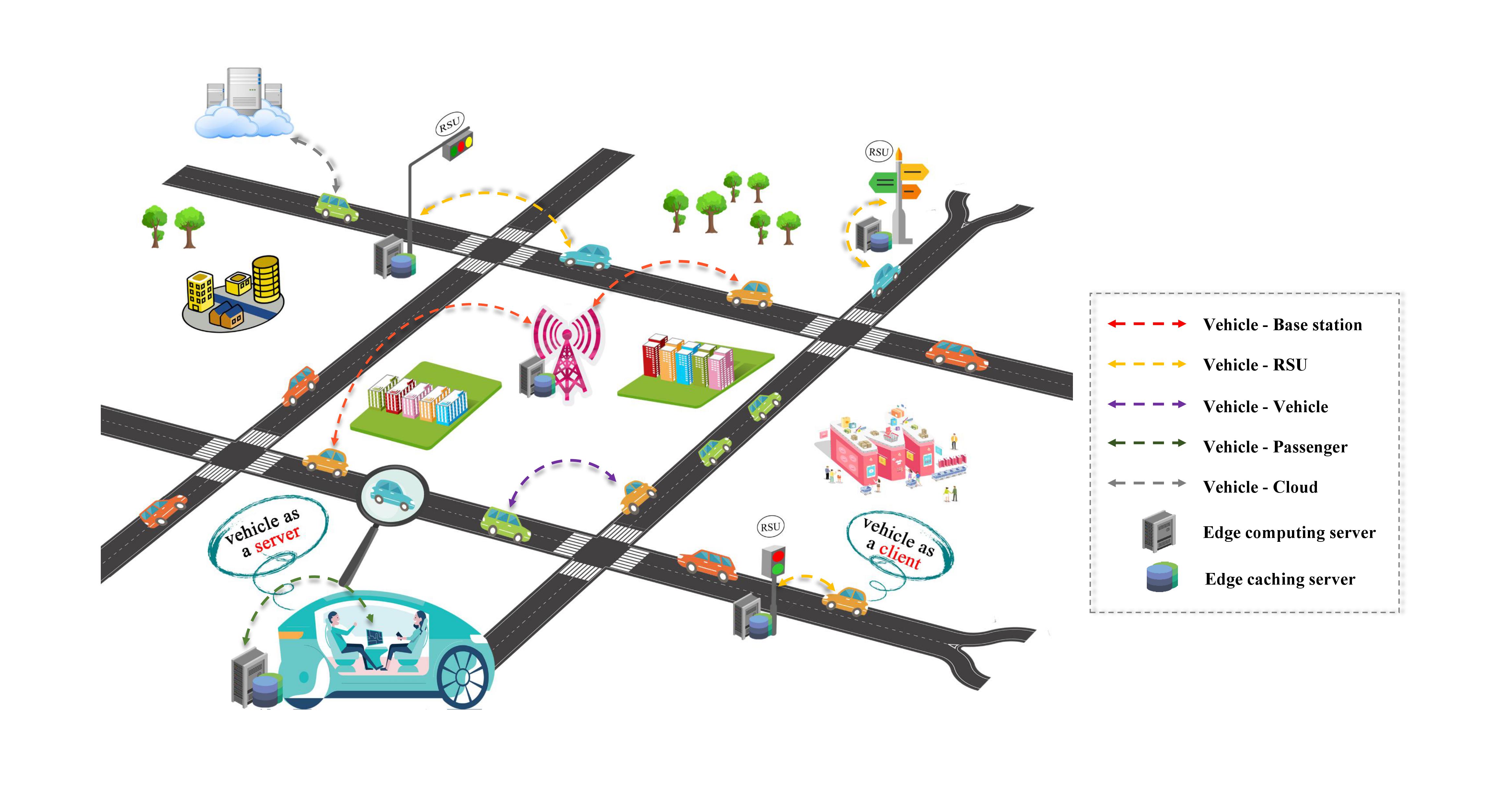}
\caption{Intelligent IoV supported by the edge information system.}
\label{fig:System}
\end{figure*}

EIS is a perfect fit for intelligent IoV. It is able to assist the key functionalities of intelligent vehicles, from data acquisition (for situational and environmental awareness), data processing (for navigation, path planning), to actuation (maneuver control), as illustrated in Fig. \ref{fig:System}. Processing the data at the network edge can save a significant amount of communication bandwidth, and also satisfy the low latency requirement for mission-critical tasks. Content in IoV is typically with high \emph{spatial locality}, e.g., road conditions and map information are mainly used locally, and \emph{temporal locality}, e.g., traffic conditions in the morning will be of little relevance for the evening. Moreover, vehicles are only interested in content itself, not its provenance. These key features make cache-assisted content-centric dissemination and delivery highly effective for IoV \cite{amadeo2016information,coutinho2018design}. Furthermore, some content is costly to download directly from Internet, and is requested repeatedly by multiple users, e.g., map download and Tesla's over-the-air software update. In such cases, cache-assisted delivery can significantly save communication overheads. On the other hand, with big sensing data, intelligent vehicles are facing tremendous computation burdens. For example, computational capability remains the bottleneck that prevents vehicles from benefiting from the high system accuracy enabled by higher resolution cameras. Specifically, the convolution tasks in the powerful Convolutional Neural Network (CNN) \cite{LeCBen15} for vision-based perception and the feature extraction tasks for vision-based localization are highly complicated \cite{LiuTan17}.
Offloading such computation-intensive tasks to the proximate MEC servers will enable powerful machine learning methods to assist key tasks of intelligent vehicles \cite{ZhaMao17,5GAA17}.

While this paper focuses on EIS, intelligent IoV shall leverage different available information processing platforms:
\begin{itemize}
\item \textbf{Onboard processing} is used for highly latency-sensitive tasks, such as real-time decision making for vehicle control, and for preprocessing sensing data to reduce communication bandwidth;
\item \textbf{Edge severs} are for latency-sensitive and computation-intensive tasks, such as localization and mapping, and for aggregating and storing local information, such as the area's high-definition map;
\item \textbf{Cloud computing} is for training powerful deep learning models with massive datasets, acts as a non-real-time aggregator for wide-area information, and stores valuable historic data for continuous learning.
\end{itemize}

EIS shall play a vital and unique role in the information infrastructure for intelligent IoV. There has been no systematic survey covering this important topic, especially its applications to key tasks for intelligent IoV. We intend to fill the gap, with a comprehensive and in-depth introduction of EIS for intelligent IoV.

\subsection{Paper Outline}
\begin{figure*}[t]
\center
\includegraphics[scale = 0.5]{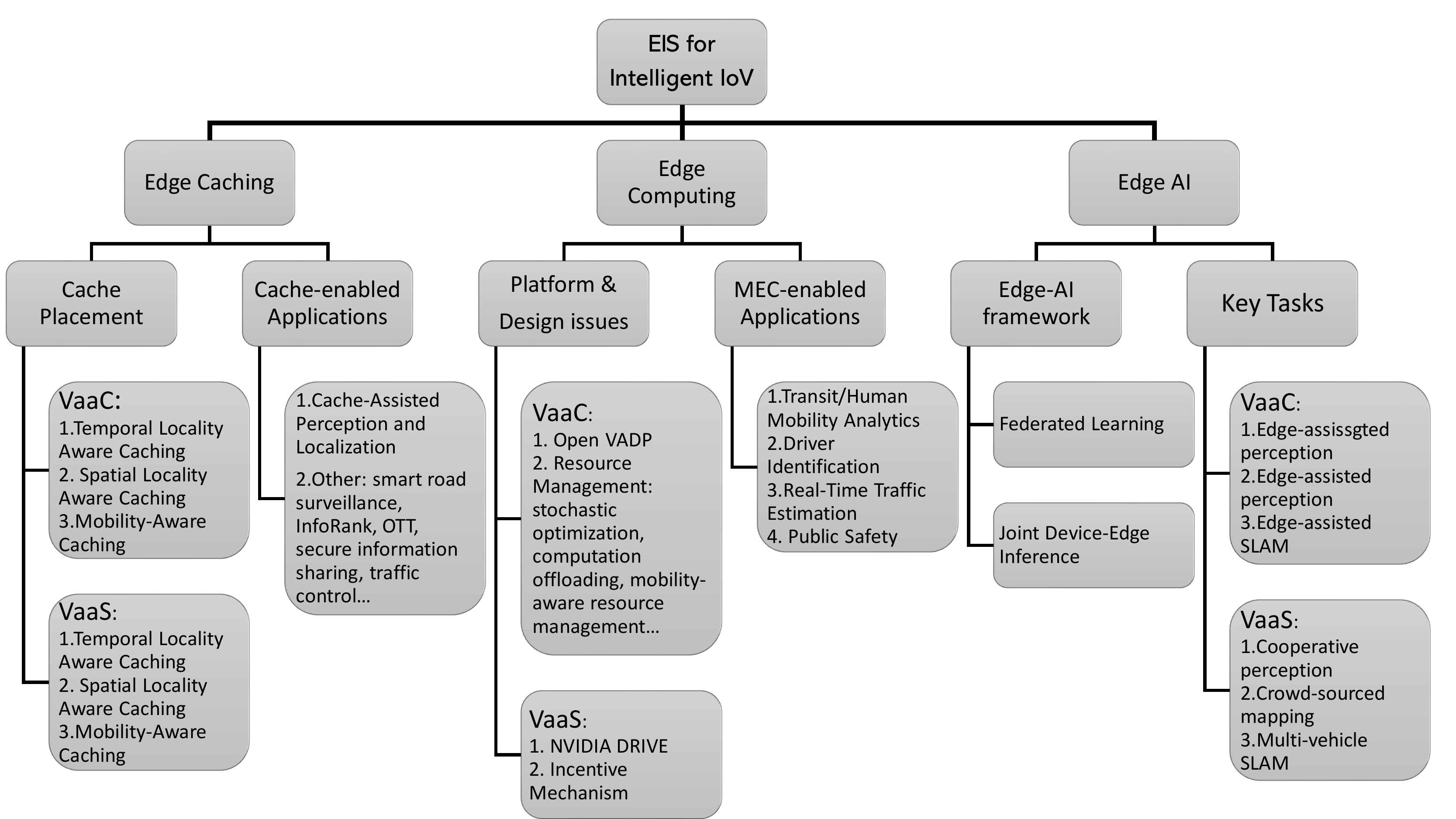}
\caption{The structure of the paper.}
\label{fig:Structure}
\end{figure*}

This article surveys recent developments in EIS for intelligent IoV, spanning edge caching, edge computing, and edge AI. Section \ref{EIS} presents a general introduction of EIS for IoV, as well as key tasks for intelligent vehicles that will be used as application examples. Section \ref{Caching} introduces edge caching systems for IoV, focusing on cache placement and delivery, as well as cache-enabled applications for intelligent IoV. Section \ref{MEC} introduces MEC platforms for IoV, key design problems, and MEC-enabled applications. Section \ref{EdgeAI} describes edge AI frameworks, and illustrates how EIS helps key tasks in intelligent IoV. Finally, Section \ref{conclusion} concludes the paper. For ease of reference, acronyms used in the paper are listed in Table \ref{tab:Acronym}, and the paper structure is shown in Fig. \ref{fig:Structure}.

\begin{table}
    \centering\small
    \caption{\label{tab:Acronym}Used acronyms.}
\begin{tabular}{|c|l|}
    \hline
    5G & The Fifth Generation \\
    \hline
    ADAS & Advanced Driver-Assistance Systems\\
    \hline
    AI & Artificial Intelligence\\
    \hline
    BS & Base station\\
    \hline
    CNN & Convolutional Neural Network\\
    \hline
    DNN & Deep Neural Network\\
    \hline
    DSRC & Dedicated Short Range Communications\\
    \hline
    EIS & Edge Information System \\
    \hline
    GNSS & Global Navigation Satellite Systems\\
    \hline
    GPS & Global Positioning System\\
    \hline
    GPU & Graphics Processing Unit\\
    \hline
    HD Map &  High-Definition Map\\
    \hline
    IoT & Internet of Things\\
    \hline
    IoV & Internet of Vehicles\\
    \hline
    KPI & Key Performance Indications\\
    \hline
    LiDAR & Light Detection And Ranging sensors\\
    \hline
    LSTM & Long Short-Term Memory\\
    \hline
    MEC & Mobile Edge Computing \\
    \hline
    MLP & Multi-Layer Perception\\
    \hline
    RSU & Roadside Unit\\
    \hline
    SLAM & Simultaneous Localization and Mapping\\
    \hline
    SSD & Solid-State Drive\\
    \hline
    TPU & Tensor Processing Unit\\
    \hline
    V2I & Vehicle-to-Infrastructure communications\\
    \hline
    V2V & Vehicle-to-Vehicle communications\\
    \hline
    V2X & Vehicle-to-Everything communications\\
    \hline
    VaaC & Vehicle as a Client \\
    \hline
    VaaS & Vehicle as a Server \\
    \hline
    VANET & Vehicular Ad-hoc Networks\\
    \hline
\end{tabular}
\end{table}

\section{The Edge Information System for IoV}
\label{EIS}

This section first introduces EIS for intelligent IoV, considering two scenarios: Vehicle as a Client (VaaC) and Vehicle as a Server (VaaS). Then, three key tasks for intelligent vehicles are presented as key application cases, including perception, high-definition (HD) mapping, and Simultaneous Localization and Mapping (SLAM). Fig. \ref{fig:AV} and Table \ref{tableI} provide a quick reference.

\subsection{The Edge Information System}

EIS helps to acquire, aggregate, and process data for intelligent vehicles, with the help of various edge resources and vehicular communications. The objective is to improve driving safety and efficiency, and enhance the capability of each vehicle. Within IoV, it acts as an intermediary platform between onboard processors and the remote Cloud data center.

As shown in Fig. \ref{fig:System}, an EIS contains the following main components.

\textbf{Edge servers}. These are computer servers deployed at BSs or roadside units (RSUs), equipped with storage units, such as SSD, and computing units, such as GPUs or Edge TPUs (Tensor Processing Units) \cite{TPU}. They are connected to the backbone network with high-capacity links, and are capable to communicate with vehicles within their coverage ranges via V2I communications. BS servers have larger coverage ranges, and thus have better capability for mobility management. On the other hand, RSU servers are closer to vehicles, and thus can support lower latency. Both types of edge servers are aware of the local environment, can collect and process data from vehicles passing by, e.g., to update HD maps or monitor traffics, and disseminate content such as road condition, traffic information, and HD maps.

\textbf{Vehicles}. Equipped with various sensors, communication modules, and onboard units with computing and storage capabilities, intelligent vehicles are powerful nodes \cite{AbdHas15}. The onboard computing stack must simultaneously achieve high performance, consume minimal power, have low thermal dissipation, while at an acceptable cost \cite{LiuTan17}. Vehicles are able to communicate with each other via V2V communications, or with the RSUs via V2I communications. The throughput of V2I communications is typically larger than that of V2V communications \cite{YuBai11}.

\textbf{User devices}. User devices are of a variety of types, e.g., passengers' smartphones and wearable devices. They are typically with limited computing power, storage space, and power supply. High end devices may be equipped with AI chips, but the computing power is still limited compared with onboard or edge server processors. These devices will generate user specific data that can be used to improve driving experience and safety.

We shall consider two different scenarios, depending on the role of vehicles, as specified below.

\subsubsection{Vehicle as a Client (VaaC)}
First, vehicles may act as clients to access the edge resources at the RSUs or BSs. The key idea is to \emph{co-locate the data acquisition and processing}. Edge servers act as anchor nodes for data acquisition, and then process data for local applications. For example, they can collect mapping data from vehicles passing by to build and update HD maps, and can actively monitor the road condition and traffic in the local area. Such applications are highly relevant for IoV. While the data generated by one vehicle may be limited, each edge server can accumulate a large amount of data from many vehicles, which is essential for training powerful machine learning models and vehicle/traffic data analytics.

\subsubsection{Vehicle as a Server (VaaS)}
Vehicles can also work as mobile service providers for vehicle passengers, third-party recipients, and other vehicles. This can be used to improve user experience, e.g., to enable personalized driving experience via driver identification, and to enable rich infotainment applications. Compared with edge server based approaches, VaaS suffers less from mobility. Moreover, it also allows cooperation among neighboring vehicles, e.g., to enable cooperative perception and cooperative driving, which will lead to connected intelligence by enhancing the capability of individual vehicles.

\subsection{Key Tasks of Intelligent Vehicles}

\begin{figure*}[t]
\center
\includegraphics[scale = 0.4]{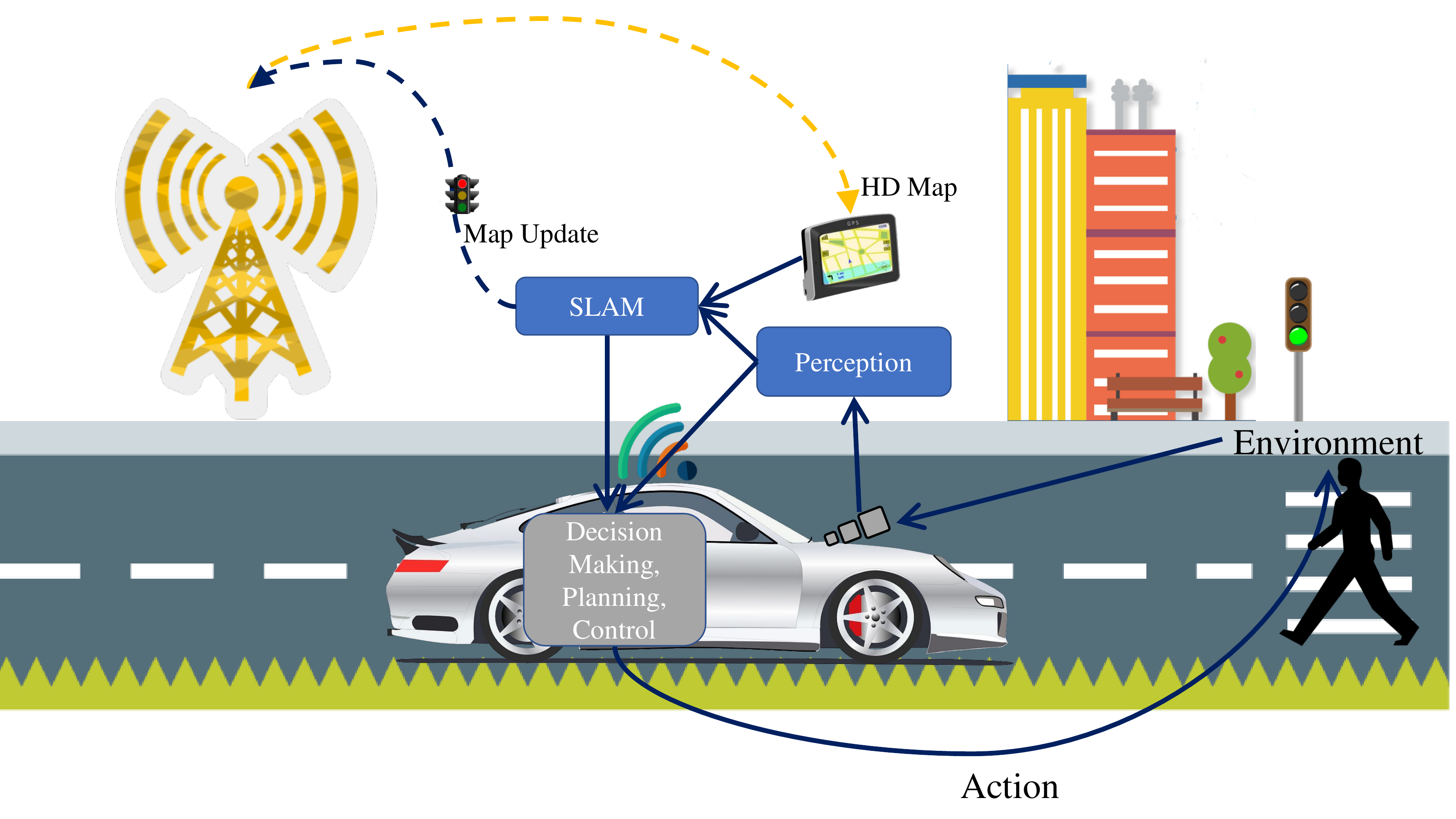}
\caption{An illustration of key tasks for intelligent vehicles, and how they act together for autonomous driving.}
\label{fig:AV}
\end{figure*}

\begin{table*}[!t]\footnotesize
        \caption{Key Tasks for Intelligent Vehicles.}
        \label{tableI}
        \centering
        \begin{tabular}{|p{1.3cm}|p{4cm}|p{4cm}|p{6.5cm}|}
                \hline
                \textbf{Task} & \textbf{Function} & \textbf{Challenges} & \textbf{Edge-Assisted Approaches} \\
                \hhline{|=|=|=|=|}
                Perception & Estimate the environment model with on-board sensors, e.g., object detection and tracking, lane detection. & High computation intensity for deep learning models; limited perception range. & Cooperative perception; edge-assisted deep learning for vision-based localization; edge-assisted multi-sensor fusion. \\
                \hline
                HD Mapping & For three-dimensional representation of all crucial aspects of a roadway. & Intensive works for building and updating maps; large storage space; significant communication overhead for dissemination. & Edge-assisted map building and update; caching-assisted data aggregation and map dissemination; multi-vehicle crowdsourced mapping. \\
                \hline
                SLAM & Simultaneous estimation of the location of a vehicle and the construction of the map. & High computational intensity; real-time execution. & Edge-assisted SLAM, multi-vehicle SLAM. \\
                \hline
        \end{tabular}
        \vspace*{-15pt}
\end{table*}

To introduce EIS for intelligent IoV, we shall focus on a few key tasks of intelligent vehicles.
The foundation of intelligent vehicles is the ability to understand the environment. Different on-board sensors, including ultrasonic sensors, Radar, LiDAR, cameras, and GNSS, are employed for different perception tasks, e.g., object detection/tracking, traffic sign detection/classification, lane detection, etc. The \emph{a priori} knowledge, e.g., \emph{a priori} maps, is also exploited. Based on the sensing data and perception outputs, localization and mapping algorithms are applied to calculate the global and local location of the vehicle and map the environment. The results from these tasks are then used for other functions, including decision-making, planning, and vehicle control, as illustrated in Fig. \ref{fig:AV}. In this paper, we shall focus on perception, HD mapping and SLAM as the main tasks for intelligent vehicles, and illustrate how EIS can help improve their efficiency and accuracy. In this subsection, a brief introduction is provided for each task, and the connection with EIS is established, as summarized in Table \ref{tableI}. More details on edge-assisted approaches for performance enhancement will be provided in later sections.

\subsubsection{Perception}
There are various kinds of on-board sensors, with different features and serving for different perception tasks. Cameras are used for object detection/classification, e.g., via the powerful CNN \cite{LeCBen15}, and can also be used for vision-based localization \cite{WolEus14}; LiDAR is applied for 3D mapping and localization \cite{ThrBur05}; Radar and Sonar are used for obstacle detection. Each type of sensors has its limitations. Cameras and stereovision are computationally expensive compared with active sensors such as LiDAR and Radar, while LiDAR and Radar are poor in classification and poor for very near ($<2$ m) measurement, and sonar has poor angular resolution.

Considering budget and features of different technologies, developers may choose different combinations of sensors for an intelligent vehicle. For example, Waymo is using the LiDAR-based technology, while Mobileye and Tesla are relying on cameras and sensors \cite{Bar16}. Currently, the limitations and high costs of available on-board sensors are one main reason that commercial vehicles only achieve Level 1 to Level 2 automation \cite{BruObr18}. The perception of intelligent vehicles faces a few main challenges, such as perception in poor weather and lighting conditions, or in complex urban environments, and limited perception ranges. Techniques such as sensor fusion can be used to compensate for shortcomings of individual sensors, by exploiting sensing data from different sensors \cite{BrsRah16,DarMan17,GruBel16}. However, this will significantly increase the on-board computation.

By providing additional proximate computation and storage resources, EIS is capable of enhancing the perception capability. It can help to improve the sensing accuracy of cameras and stereovision, e.g., via powerful deep learning techniques, and enable sophisticated multi-sensor fusion, by offloading computation-intensive sub-tasks to edge servers.  Furthermore, by sharing on-board sensing and computing power, assisted by V2V and V2I communications and coordinated by edge servers, cooperative perception can significantly improve the sensing robustness and accuracy, and extend the perception range \cite{KimLiu15}.

\subsubsection{HD Mapping}
Mapping is fundamental for any mobile robotics applications, and it is especially important for autonomous driving. Recently, HD mapping has received lots of attention. An HD map models the surface of the road to an accuracy of 10$\sim$20 cm. It contains the three-dimensional representation of all crucial aspects of a roadway, e.g., slope and curvature, lane marking types, and roadside objects. Localization with HD maps can achieve centimetre-level precision. It uses onboard sensors to compare an vehicle's perceived environment with the corresponding HD map. This can overcome the limitations of GNSS based methods (e.g., GPS), including low positioning accuracy and the varying availability. It is expected that HD map-based localization will be a common approach in Level 4 and Level 5 autonomous driving systems. HD map also helps to resolve two main issues of perception based on on-board sensors \cite{KimCho18}:
\begin{itemize}
\item \textbf{The absence of integrity}: The integrity of the information acquired from internal sensors is not guaranteed, as the sensors are vulnerable to environmental noises and may be limited in their capability due to cost concerns.
\item \textbf{The limited perception range}: Sensors have limited perception ranges, and they cannot measure target objects located outside of its Field of View (FoV) or target objects occluded by other obstacles.
\end{itemize}

Great efforts from the industry have been put forward to build HD maps, e.g., TomTom \cite{TomTom} and HERE \cite{HERE}. Nevertheless, there are practical implementation difficulties. To create an HD map is time-consuming. To generate the HD map, specialized mapping vehicles equipped with Mobile Mapping System (MMS) are needed, and the whole process involves three procedures: data acquisition to acquire the mapping data, data accumulation to accumulate features acquired by the mapping vehicles, and data confirmation to manually refine and confirm the map. Furthermore, HD maps are dynamic and need to be updated with timely changes \cite{AlcSte18}, for which we need to collect fresh HD data timely and detect changes from the fresh data \cite{Jiao18}. Some HD map suppliers work with automakers to get fresh map data from intelligent vehicles, but this will substantially increase the on-board processing burden of vehicles \cite{SeiHu16}.

Furthermore, the HD map data is of very large size, due to the high precision, and the rich geometric information and semantics \cite{Jiao18}. This causes difficulties to transmit and store the HD maps. They are usually being served from the Cloud, with a few nearby small areas of the map downloaded to the vehicle. The amount of data to download for a 3D HD map with a centimetre level accuracy can reach 3$\sim$4 GB/km. This not only introduces latency for the data download from the Cloud, but also causes a heavy burden on the backbone network.


With the inherent geographic locality, EIS will play an important role in HD mapping for intelligent IoV. Different edge-assisted approaches can be employed. Edge caching can help HD map dissemination, as well as mapping data aggregation. Edge computing can assist map building and map change detection, by exploiting locally cached data. Edge servers can also coordinate vehicles driving through the area for crowdsourced mapping. In this way, by keeping and processing the data locally, and constructing the map where it is needed, more efficient HD mapping can be achieved.


\subsubsection{Simultaneous Localization and Mapping (SLAM)}

Map-based localization is effective for driving along roads that do not change often. However, if drastic changes occur, the loss in the accuracy may affect driving safety. SLAM comprises the simultaneous estimation of the state of a vehicle and the construction of a map of the environment \cite{LeoDur91,DurBai06,CadCar16,BreAls17}. It does not rely heavily on \emph{a priori} information, and allows vehicles to continuously observe the environment and readily adapt to new situations. To achieve full autonomy, it is a necessity for an intelligent vehicle to be able to perform accurate SLAM within its environment \cite{DurBai06}. Moreover, redundancy is needed in order to improve safety and ensure a consistent behaviour on the road, and thus different localization methods, i.e., map-based and SLAM-based methods, should be considered. SLAM has been regarded as a key enabling technique for autonomous driving, and vehicles from the 2007 DARPA Urban Challenge have already used SLAM-based methods \cite{UrmBag07}.

While many SLAM algorithms have been developed, they are mainly for indoor, highly structured environments. Autonomous vehicles need to operate in outdoor, variably-lit, road-based environments, and thus faster and more efficient algorithms are needed. For challenges of SLAM for autonomous driving, the readers are referred to \cite{BreAls17} for a more detailed discussion. We shall mainly discuss aspects that are related to EIS. Particularly, the computation demand for SLAM will be highly intensive for autonomous driving. One hour of drive time can generate one terabyte of data, and interpreting one terabyte of collected data by means of high computing power requites two days to come up with usable navigation data \cite{SeiHu16}. Moreover, for real-time execution, latency must be lower than 10 ms, which puts high pressure for onboard computing.

While Cloud-based SLAM algorithms have been proposed to alleviate the computation burden of vehicles, the propagation latency will not meet the real-time execution requirement. The edge computing platform can resolve this difficulty, and it can help handle part of the computation-intensive subroutines \cite{RiaCiv14}. Multi-vehicle SLAM with cooperation among vehicles can also help to improve the performance of SLAM \cite{BreAls17}. More discussions will be provided in Section \ref{EdgeAI}.


\section{Edge Caching for Intelligent IoV}
\label{Caching}

Edge caching has been adopted in IoV to assist content delivery by storing or prefetching content at edge servers \cite{amadeo2016information}. The design of caching algorithms in IoV is more challenging compared with traditional networks, caused by the high mobility of vehicles, frequently-changing content requirements, harsh communication environments, etc. In this section, we survey the existing research on caching for IoV with vehicles taking different roles, i.e., \textit{Vehicle as a Client} and \textit{Vehicle as a Server}. In the former scenario, vehicles act as content consumers in cache-enabled IoV to access the desired content from edge servers. In the latter scenario, vehicles also act as content providers by caching content at their storage units. We focus on cache content placement, i.e., to determine which content should be cached, and divide existing studies into three categories: a) \emph{Temporal locality-aware caching}, i.e., accounting for the temporal variation of the importance/popularity of contents; b) \emph{Spatial locality-aware caching}, i.e., considering different importance/popularity of the same content in different regions; c) \emph{Mobility-aware caching}, i.e., alleviating the effect of vehicle mobility on content caching and delivery.


\subsection{Vehicle as a Client}
\label{VehicleClient}

With the wide deployment of edge servers, vehicles that drive through their coverage areas can receive timely content delivery services. When designing caching systems for IoV, key features of typical scenarios should be considered. For highway scenarios, due to the sparse distribution of edge servers, the intermittent connections with requested vehicles become the main limitation of caching services. For urban roads, the high density of vehicles results in diverse content requirements, which significantly challenges caching policy making. The representative studies from traditional IoV to intelligent IoV are discussed below in detail.

\subsubsection{Temporal Locality-Aware Caching}

The temporal locality feature of vehicular services plays a vital role when satisfying user requests. In edge caching systems, temporal locality includes two aspects: freshness of cached content and the temporal variation of user requests. By considering the characteristics of communications between RSUs and vehicles, the temporal data dissemination problem was shown in \cite{liu2014temporal} as an NP-hard problem. Authors then exploited a heuristic scheduling algorithm according to the requirements of user requests (e.g., the time bound), in order to improve the request service chance. Focusing on temporal information services in IoV, a distributed edge caching mechanism was proposed in \cite{dai2016towards} based on the cooperation of RSUs and vehicles, in order to optimize both the temporal data and real-time requests. Through a multi-objective optimization algorithm, the trade-off between service quality and service ratio was obtained.

\subsubsection{Spatial Locality-Aware Caching}
In vehicular environments, some information, e.g., traffic information, is associated with the location of vehicles. Therefore, the freshness of content in IoV may vary in different road segments. In \cite{ndikumana2018deep}, a deep learning based caching scheme was developed to optimize caching decisions for intelligent IoV, aiming at reducing the delivery delay of entertainment content. In this scheme, the content is cached at the edge servers in some specific areas by detecting the ages and genders of passengers with a CNN and predicting a proper content with a Multi-Layer Perception (MLP). The vehicles then determine which content could be accessed from edge servers based on a k-means algorithm and binary classification. Similarly, a data service scheme with the assistance of caching was developed in \cite{abdelhamid2015road} by considering location-based vehicular services. A concept of road caching spot was applied to meet the service requirements and reduce the caching cost.

\subsubsection{Mobility-Aware Caching}
To deliver large-size content (e.g., videos, music, and HD maps) to moving vehicles is challenging because of limited network capacities and intermittent connections. To minimize the downloading time of vehicles, the problem of how to place large-size content at the edge servers was investigated in \cite{ding2015roadside}. Three algorithms were developed to alleviate the impact of the mobility of vehicles on caching performance. In \cite{hou2018q}, a caching strategy was proposed to minimize the caching service delay in a multi-access EIS. Specifically, the mobility of vehicles was predicted by a Long Short-Term Memory (LSTM) network over a time sequence. Based on the prediction, a deep reinforcement learning algorithm was then used to develop a proactive caching strategy. To tackle the mobility of vehicles and meet the service deadline, a new research direction was to integrate edge caching and computing in EIS. In this case, how to effectively allocate the limited resources is an important problem. Focusing on resource allocation in the integrated architecture, two joint optimization models were formulated in \cite{tan2018mobility, he2018integrated} to determine the optimal caching and computing decisions, which were then solved by deep reinforcement learning based methods.


\subsection{Vehicle as a Server}
\label{VehicleServer}

Edge server based caching is limited by the coverage range and unreliable connections with vehicles. Caching content on moving vehicles serves as a good complement. By exploiting the mobility of vehicles, edge caching can provide more cost-effective and utility-enhanced services. Existing studies related with this direction are presented as follows.

\subsubsection{Temporal Locality-Aware Caching}

For vehicle caching, the temporal locality of content impacts not only caching services, but also the implementation of other functions on vehicles due to the limited onboard storage resources. Therefore, to determine how long the content will be cached is a vital problem. In \cite{fiore2011caching}, edge caching according to the content size was proposed. A new caching method, named Hamlet, was proposed to generate content diversity among adjacent nodes by determining caching updating frequency for large-sized and small-sized content. Based on the proposal,  users could receive different contents from nearby caching nodes in a short time, which improves the caching efficiency. Likewise, Malandrino et al. \cite{malandrino2012content} focused on the freshness of the content. By studying the impact of the number of caching nodes on users, the authors optimized both the content freshness and user downloading experiences.

\subsubsection{Spatial Locality-Aware Caching}

The vehicle as a server mode will improve the caching performance for the location based services, thanks to the flexible mobility of vehicles and multi-hop data transmissions. Caching services in hot spot areas were investigated in \cite{yao2018cooperative}. Specifically, the urban areas were divided into many hot regions based on the dynamic mobility and density of vehicles. The driving traces of vehicles in the near future were predicted by partial matching based on the history data. By incorporating the vehicles that visit those hot regions frequently in a cooperative caching scheme, the optimal utility of caching services could be obtained. To mitigate the impact of mobility and communications vulnerability on caching services, a dynamic relay strategy for in-vehicle caching was developed in \cite{hu2019vehicle}. With the caching scheme and inter-vehicle communications, the survival of content in hot regions could be maintained. Besides, erasure codes were adopted to reduce caching redundancy with the existence of volatile V2V links.

\subsubsection{Mobility-Aware Caching}
The predictable mobility of vehicles can be exploited to improve the efficiency of cache-assisted content delivery. Mobility-aware caching in conventional device-to-device networks has been well studied, e.g., \cite{WanPen16,WanZha17,WanZha18}, and such methods have recently been extended to vehicular networks. A new type of caching services was explored in \cite{zhang2019mobility, vigneri2018low}, where the content cached in vehicles could be requested by moving or static users within the communication range. In this scenario, the relationship between caching vehicles and moving users was the key to design the caching policy. A $2$-D Markov process was proposed in \cite{zhang2019mobility} to model the interactions of caching vehicles and moving users, in order to determine the network availability of mobile users. Vigneri et al. \cite{vigneri2018low} analyzed the playout buffer at each user device by queueing theory, and optimized content placement on the moving vehicles.

\subsubsection{Parked Vehicles for Caching}
There are proposals to utilize parked vehicles as cache servers due to the convenience in connecting with the edge servers and moving vehicles \cite{liu2011pva, su2017game, elsayed2018proactive}. In these studies, parked vehicles play a similar role as fixed edge servers. Liu et al. \cite{liu2011pva} unfolded the journey of caching at parked vehicles by analyzing the connectivity of parked vehicles with clients. Su et al. \cite{su2017game} further developed a competition and cooperation mechanism for caching at parked vehicles to encourage parked vehicles to join in the caching system. Recently, a proactive caching method was proposed in \cite{elsayed2018proactive}, which caches social media content in parked vehicles. This is motivated by the fact that social media content is becoming one of the main contributors to the network burden of the cellular infrastructure.

\subsection{Cache-Enabled Applications}

Besides typical content sharing and delivery services, there have been great interests in developing new applications enabled by edge cache servers. This subsection first introduces cache-assisted perception and localization, followed by other applications in IoV and intelligent transportation systems.

\begin{table*}[t]
	\centering
	\caption{Use Cases of Cache-Assisted Perception and Localization.}
	\label{UseCasesinIoV}
	\scalebox{1}{
		\begin{tabular}{|p{2.5cm}|p{5.5cm}|p{3.2cm}|p{4cm}|}
			\hline
			\textbf{Use Cases} & \textbf{Functions} & \textbf{KPIs} & \textbf{Cached Content}  \\
              \hhline{|=|=|=|=|}
			Autonomous Overtake & Guarantee safety distance between the overtaking vehicle and oncoming vehicle on two-way roads.& \makecell[tl]{E2E delay: $10$ ms, \\Reliability: $10^{-5}$, \\Positioning accuracy: $30$ cm.} & Road information and vehicle intention. \\
			\hline
			{Cooperative Collision Avoidance} & Identify collision risks (such as in intersections) and inform vehicles in advance when traffic control mechanism fails. & \makecell[tl]{E2E delay: $10$ ms, \\Reliability: $10^{-3}$, \\Positioning accuracy: $30$ cm.} & Vehicle driving trajectories. \\
			\hline
			See-Through & Obtain the view of blind spots obstructed by nearby vehicles in AR/VR. & \makecell[tl]{E2E delay: $50$ ms, \\Data rate: $10$ Mbps.} & Surveillance video/image for road segments. \\
			\hline
			Bird's Eye View & Provide the surveillance content of special road segments for approaching vehicles. &\makecell[tl]{E2E delay: $50$ ms, \\Data rate: $40$ Mbps.} & Surveillance video/image for road segments. \\
			\hline
			High Density Platooning & Form multiple vehicles into a linear chain by cooperative driving. & \makecell[tl]{E2E delay: $10$ ms, \\Reliability: $10^{-5}$, \\Positioning accuracy: $30$ cm.} & Vehicle driving behaviors, situational awareness, lane changing information. \\
			\hline
			VRU Discovery & Detect vulnerable users by exchanging the localization information of vehicles and users. &Positioning accuracy: $10$ cm. & Localization information of vehicles and pedestrians. \\
			\hline
	\end{tabular}}
\end{table*}

\subsubsection{Cache-Assisted Perception and Localization}
As reported in 5G Automotive Vision (5GAV) \cite{5GAV}, new use cases of edge caching will emerge in the era of intelligent IoV. In this part, we focus on applying content caching for two key tasks of intelligent vehicles, namely, perception and localization. Cache-assisted perception includes such functions as autonomous overtake, cooperative collision avoidance, see-through, and bird's eye view, in which edge caching provides perception content for vehicles to assist driving and improve traffic safety. On the other hand, cache-assisted localization includes vulnerable road user (VRU) discovery, in which edge caching improves the cooperation of RSUs, vehicles, and pedestrians by caching positioning information. Table \ref{UseCasesinIoV} shows the detailed description of these use cases, as well as their Key Performance Indicators (KPIs) in terms of end-to-end (E2E) delay, reliability, data rate, and positioning accuracy. The types of cached content are also shown.

\subsubsection{Other Applications}
New applications enabled by edge caching keep emerging in IoV, some of which are introduced below.

\textbf{Smart Road Surveillance}: A virtual resource platform for IoV was elaborated in \cite{LeeGer16} to support real-time video surveillance of traffic events. The edge services in the platform provide flexible and fast access for moving vehicles. With cooperation among edge servers and vehicles, the video of a temporary traffic event (e.g., a minor accident ahead) could be collected and cached in edge servers, which is then delivered to the oncoming vehicles.

\textbf{InfoRank}: For efficient urban sensing, Khan et al. \cite{khan2016autonomous} developed an information based ranking (InfoRank) algorithm. This algorithm selects and ranks a part of intelligent vehicles to undertake urban sensing tasks. The vicinity monitoring of those vehicles thus could be completed with a small cost. In the algorithm, vehicles act as data cache servers to store the sensing data, in order to alleviate the burden of edge servers.

\textbf{OTT}: A new Over-The-Top (OTT) content prefetching system was designed in  \cite{zhao2018mobility} by implementing an edge caching mechanism. The connections of vehicles and RSUs are predicted based on a real-world testbed. A content popularity estimation scheme is also developed to estimate the content requests of users. After that, the requested content of users is proactively prefetched at edge servers.

\textbf{Secure Information Sharing}: Data sharing is an efficient way to reduce the data loss caused by unreliable sensor systems and to overcome the limited sensing range in autonomous vehicles. Data security thus becomes an important task, and Chowdhury et al. \cite{chowdhury2017secure} designed a secure information sharing system for autonomous vehicles. The system aims to improve the data security in two scenarios, including false data dissemination, and vehicle tracking.

\textbf{Traffic Control}: To analyze the impact of edge caching on traffic control, an edge caching based transportation  control scheme was developed in \cite{liu2018traffic}. Traditionally, it is difficult to obtain the optimal state of the transportation system since drivers are selfish. As such, the optimal state of transportation networks and user equilibrium are contradictory. A communication cost model for cache-enabled vehicles was proposed to uncover the relationship between the user equilibrium and system optimal state. With the proposal, transportation networks could be optimized from a communication aspect with the assistance of edge caching.


\section{Edge Computing for Intelligent IoV}
\label{MEC}

With edge servers in close proximity to mobile users, MEC brings a number of important benefits, including ultra-low latency, reduced mobile energy consumption, and enhanced privacy and security.  This section introduces edge computing platforms for intelligent IoV. Available hardware platforms are introduced first, and key design problems are then discussed. Examples of MEC-enabled applications in IoV are also presented.

\subsection{Vehicle as a Client}
We first consider the VaaC scenario, where vehicles act as clients to access the computation resources at edge servers.

\subsubsection{Edge Computing Platforms for Intelligent IoV}
The concept of MEC was first proposed by ETSI in 2014 \cite{hu2015mobile}, and it has been regarded as a key component of the upcoming 5G network. Edge computing servers, equipped with GPUs or Edge TPUs \cite{TPU}, can be deployed at different edge nodes, and we mainly consider BSs and RSUs.  The implementation of edge servers at BSs relies on several key techniques of 5G networks, such as the network virtualization architecture, network function virtualization, and virtual machine (VM) \cite{hu2015mobile}. The virtualization layer aggregates the geographically distributed computing resources and presents them as a single resource pool for use by applications in the upper layers. Different applications share the aggregated computation resources via VMs. Meanwhile, the capability of RSUs has been improved significantly. They normally adopt powerful multi-core CPU and massive storage units. For example, the Cohda RSU\footnote{\url{http://www.cohdawireless.com/solutions/hardware/}.} MK5 has the cortex-A8 based ARMv7 Processor, which provides high flexibility for cooperative computing applications in IoV.

There have been lots of efforts in developing edge computing platforms specialized for vehicular data analytics. Open Vehicular Data Analytics Platform (OpenVDAP) \cite{ZhaWan18} is an open-source platform. It is a full-stack edge based platform, including an on-board computing/communication unit, an isolation-supported and security and privacy-preserved vehicle operation system, an edge-aware application library, as well as an optimal workload offloading and scheduling strategy. To evaluate different edge computing platforms, CAVBench was proposed in \cite{WanLiu18}. It is a benchmark suite for edge computing in connected and autonomous vehicles, including six applications: SLAM, objective detection, object tracking, battery diagnostics, speech recognition, and edge video analysis.

Server placement of conventional MEC systems, i.e., to install edge computing servers among the available sites, has been investigated in \cite{XuLia16,CesPre17}. Recently, the optimal deployment and dimensioning of MEC-based IoV infrastructure was investigated in \cite{YuLin19}. Two different modes were compared: The coupling model (CRF), where the RSU and the edge server are co-located, and the decoupling model (DRF), where the RSU and the edge server are placed in diverse areas. It was found that the DRF mode is more cost-effective, flexible, economical, and practical.

\subsubsection{Resource Management}
Effective resource allocation is essential for computation offloading in MEC. Such allocation faces a few critical issues. First, the stochastic nature of wireless channels and task arrivals should be considered \cite{ZhaWen13,MaoZha16}; Second, the limited radio and computing resources are shared by multiple users, both of which will affect the computation latency \cite{YouHua17,MaoZha17}; Finally, the mobility of vehicles will affect the task offloading and result feedback \cite{SunZho17}.

Stochastic optimization for resource management was considered in \cite{DuYu19}. It minimizes the cost of both vehicles and the MEC server, by jointly optimizing the offloading decision and local CPU frequency on the vehicle side, and the radio resource allocation and server provisioning on the server side. A contextual architecture was proposed in \cite{LamAgr19} for MEC in vehicular networks, which evaluates available resources in real-time and assigns the most logical and feasible resource to tasks.

Computation offloading decision making among multiple vehicles was investigated in \cite{LiuWan18} by formulating it as a multi-user computation offloading game. The existence of Nash equilibrium (NE) of the game was proved, and a distributed computation offloading algorithm was proposed to compute the equilibrium. If too many tasks were offloaded to the same edge server, the performance gain will be degraded. Load balancing among edge servers when designing the offloading decision was investigated in \cite{DaiXu19}. A joint load balancing and offloading problem was formulated as a mixed integer non-linear programming problem to maximize the system utility. Such problems are highly challenging, and thus a low-complexity sub-optimal algorithm was developed by decoupling the problem as two sub-problems.

Mobility-aware resource management for MEC has also received lots of attention.  In \cite{SunZho17}, an online energy-aware mobility management scheme was developed, accounting for the radio handover and computation migration cost. An effective mobility-aware offloading decision algorithm was proposed in \cite{YuChe18}, by integrating mobility prediction. In \cite{OuyZho18}, performance optimization under a long-term cost budget constraint was investigated. Lyapunov optimization was applied, while the task migration cost was accounted for. Compared with the above results, the study in \cite{WanZho18} was specifically for vehicle networks. It considered mobile devices in legacy vehicles, running infotainment applications, and offloading some computation to nearby intelligent vehicles. Due to high mobility, it is not efficient to always offload to one vehicle. It thus proposed an edge server relaying scheme to better utilize computation resources on the road. To develop better mobility management methods, cooperation among multiple edge serves, and integration of BS and RSU servers will be needed.

\subsection{Vehicle as a Server}
With powerful onboard processing capabilities, vehicles can act as servers to provide computation services for passengers, or cooperate with other vehicles. For this purpose, incentive mechanisms are needed to encourage vehicles to share resources. Related studies are surveyed below.

\subsubsection{Vehicular Cloud}
Onboard computing is getting more and more powerful for intelligent vehicles \cite{LiuTan17}. In particular, different platforms have been developed for autonomous driving. For example, the NVIDIA DRIVE platform includes an in-vehicle computer (DRIVE AGX) and a complete reference architecture (DRIVE Hyperion), as well as data center-hosted simulation (DRIVE Constellation) and a deep neural network training platform (DGX) \cite{NVIIA}. DRIVE AGX is built on NVIDIA Xavier, the world's first processor designed for autonomous driving. Six types of processors work together inside Xavier: An image signal processor, a video processing unit, a programmable vision accelerator, a deep learning accelerator, a CUDA GPU, and a CPU. Together, they process nearly 40 trillion operations per second, among which 30 trillion operations are for deep learning alone.



Inspired by the success and flexibility of Cloud computing in providing on-demand resources and services, the concept of Vehicular Cloud arises, which is to leverage onboard vehicular resources, such as network connectivity, computational power, storage, and sensing capability \cite{MekIss17}. It can enable various applications, such as traffic management, urban surveillance, emergency management, etc. The high traffic mobility is a major challenge in implementing a Vehicular Cloud. By analyzing existing traffic models, it was shown in \cite{BouGra18} that Vehicular Cloud computing is technologically feasible in dynamic scenarios, e.g., highways. Given the potential applications of Vehicular Cloud, lots of studies have been carried to address its design challenges and implementation issues. Different resource management problems have been studies, including scheduling \cite{GhaMuk14}, virtual machine migration \cite{RefKan14}, and computation resource allocation \cite{ZheMen15}.

\subsubsection{Incentive Mechanisms}
To utilize onboard computation resources of intelligent vehicles to assist passenger devices or other vehicles, effective incentive mechanisms are needed. There have been many studies in incentivizing players to share resources in other domains \cite{JinSon16}, and recently extensions to intelligent vehicles have been investigated.

The RSU servers are sparsely deployed and constrained by their radio coverage. To overcome such limitation, it was proposed in \cite{SuHui18} to utilize vacant computing power at vehicles. A market mechanism was developed to incentivize nearby vehicles to contribute their computing power. Distributed task allocation algorithms for cost minimization were developed. A similar study was carried in \cite{ZhoLiu19}, where vehicles act as server nodes to help with computation tasks for the MEC server at the BS. Incentive mechanism and task assignment mechanism were investigated. Specifically, it developed a two-stage computation resource allocation and task assignment approach by combining contract theory and matching theory. In the first stage, the BS designs a contract to incentivize vehicles to share their computation resources. In the second stage, the vehicles which have signed the contract act as server nodes. Task assignment via a two-sided matching game was designed. In \cite{LiWDai19}, a market mechanism was designed for computation offloading to incentivize vehicles to share resources. A Vickrey-Clarke-Groves (VCG) based reverse auction mechanism was developed.

\subsection{MEC-Enabled Applications in IoV}
Proximate MEC servers have inspired new applications in intelligent IoV, based on in-vehicle data analytics. The following are some examples.

\subsubsection{Transit/Human Mobility Analytics} The public transit system is an important part of the public infrastructure. The knowledge of transit usage is critical to evaluate current transit routes/schedules, and to make necessary adjustments. In \cite{QiKan17}, an edge computing platform, named Trellis, was deployed on public transportation vehicles for human mobility analytics. Compared to Cloud computing platforms, Trellis is able to provide services with a lower latency, greater responsiveness, and more efficient use of network bandwidth. Such platforms have the potential to improve the efficiency and quality of public transportation systems.

\subsubsection{Driver Identification} Driver-specific applications are important for shared vehicles that are used by multiple drivers. An MEC system was built in \cite{KarJai17b} to collect and analyze in-vehicle data for driver identification. The system can be used for applications such as personalization of vehicle settings (e.g., automatically adjusting entertainment, preferred temperature, configurations to driver preferences), automated vehicle use logs, driver-dependent pay-as-you-drive insurance, unauthorized vehicle use detection, etc.

\subsubsection{Real-time Traffic Estimation} Traffic estimation is an important problem in intelligent transportation systems. Existing works rely on traffic surveillance cameras, which are not available on many roads, or GPS-based speed estimation, which only provides coarse estimates. Furthermore, considering the bandwidth and latency challenges, real-time traffic estimation should be near the edge and on the vehicles themselves. With the help of a front facing camera, an MEC-assisted automated traffic estimation framework was developed in \cite{KarJai17b} for vehicle detection, vehicle tracking, and traffic estimation. The effectiveness of the system has been tested through multiple days of roadway experiments.


\subsubsection{Public Safety} Video Analytics for Public Safety (VAPS) is an important application case for IoT. Due to limitations in budgets, size, weight, and power,
as well as the complexity of public safety operations, analytic integration and optimization for VAPS is a significant challenge. In \cite{LiuZha19}, built upon the OpenVDAP platform \cite{ZhaWan18}, an IoT-enabled public safety service, called AutoVAPS, was proposed. It integrates body-worn cameras and other public safety sensors, and consists of three layers: A data layer for data management, a model layer for edge intelligence, and an access layer for privacy-preserving data sharing and access.


\section{Edge AI for Intelligent IoV}
\label{EdgeAI}

In intelligent IoV, edge AI involves training powerful machine learning models and data analytics for key tasks of intelligent vehicles. It also coordinates multiple entities at the network edge for joint inference and decision making, including edge servers at BSs and RSUs, as well as onboard processors of different vehicles. This section first introduces two edge AI frameworks for collaborative training and joint inference. Then key use cases in intelligent IoV are illustrated, including edge-assisted perception, mapping, and SLAM.

\subsection{Edge AI Frameworks}
The major Cloud providers have extended their services to the edge by developing edge learning platforms, including Amazon's AWS Greengrass, Microsoft's Azure IoT Edge and Google's Cloud IoT Edge. These platforms allow edge devices to run machine learning models trained in the Cloud.
In this part, we introduce two edge AI frameworks to illustrate the training and inference stages, respectively.

\subsubsection{Federated Learning}
\begin{figure}[h]
\center
\includegraphics[width=\columnwidth]{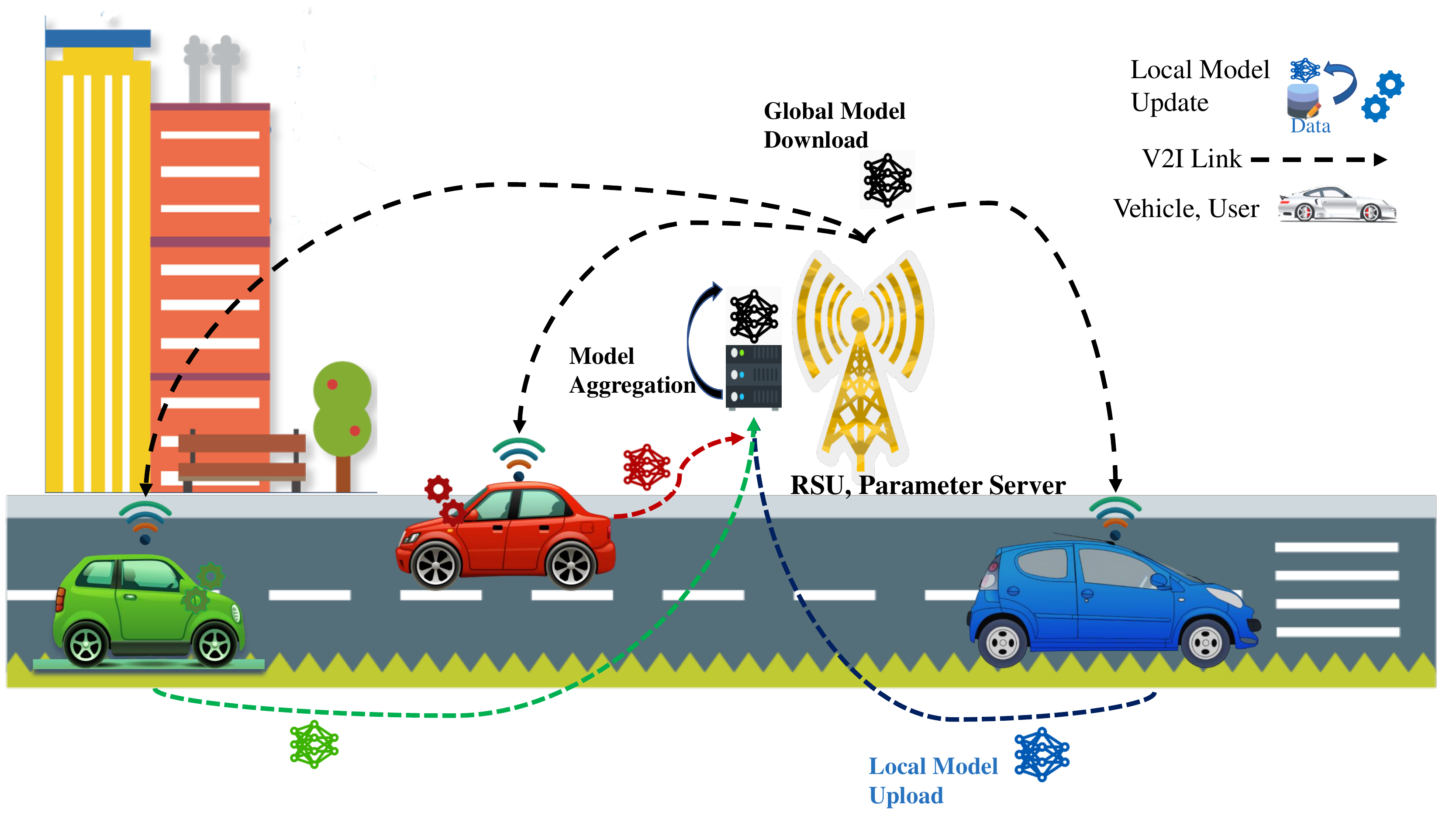}
\caption{Federated learning for collaborative privacy-preserving machine learning.}
\label{FL}
\end{figure}

One main challenge of edge AI is to train machine learning models by aggregating the large amount of data that are distributed at different edge devices, including vehicles and onboard devices. Directly moving data to a central server for training, e.g., at a Cloud server, will introduce prohibitive communication overhead. Moreover, many types of data contain personal information, and thus are privacy-sensitive. Federated learning \cite{KMR15,KonMcM16,KonMcM17,McMMoo17} is a recently proposed machine learning paradigm that allows to collaboratively train a shared model for many users without direct access to the raw data. Each user trains a local machine learning model on the local dataset, and uploads it to a server for a global model aggregation. In this way, distributed data on mobile devices can be well exploited without leak of privacy. Federated learning may help to train machine learning models for some privacy-sensitive tasks of IoV, e.g., speech recognition in the driving assistant system, and infotainment applications.

One difficulty of Cloud-based federated learning \cite{KMR15} is that the model update introduces significant communication overhead to the backbone network and long latency. The communication overhead is proportional to the machine learning model size, making it expensive to be applied to powerful deep learning models. Edge servers, acting as an intermediary between clients and the Cloud server, can be exploited to reduce the communication overhead. Specifically, we can first perform multiple local aggregations at each edge server, and then apply one global aggregation at the Cloud. As shown in a recent study \cite{LiuZha19b}, such edge-assisted hierarchical federated learning reaches a desired model performance with much less communication.

\subsubsection{Joint Device-Edge Inference}
\begin{figure}[h]
\center
\includegraphics[width=0.85\columnwidth]{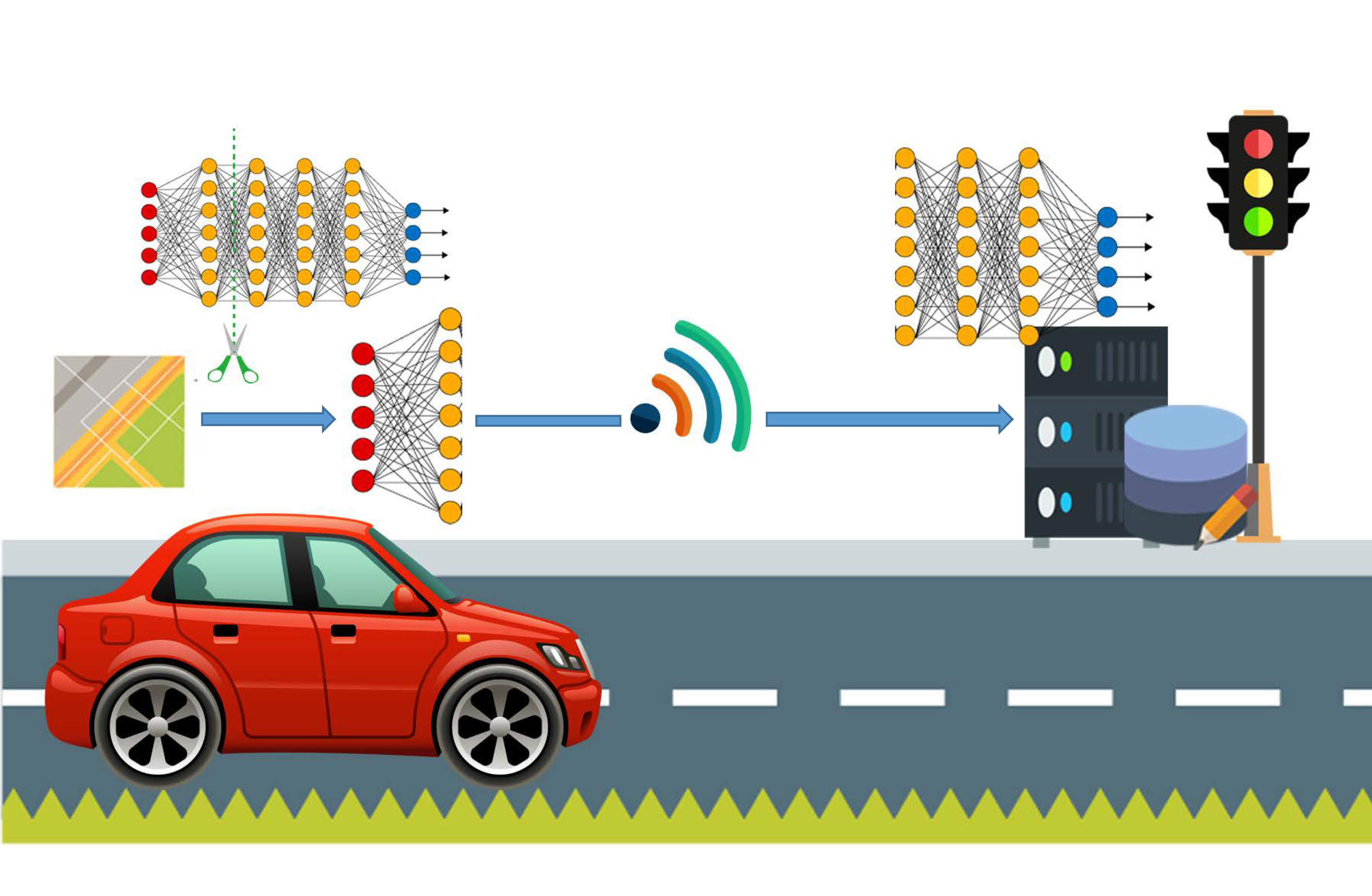}
\caption{Joint device-edge inference for intelligent vehicles.}
\label{Device-Edge-Inference}
\end{figure}

To apply computation-intensive Deep Neural Network (DNN) models, running them directly on vehicles will take too much computation resource and also consume lots of energy, while offloading them to the edge server may suffer from time-varying wireless fading channels, which leads to excessive latency when the offloading data size is large. These powerful techniques are important for vehicle perception, as shown in Section \ref{EIS}, and thus how to exploit the edge resources for efficient execution is of critical importance. Recently, joint device-edge inference has been proposed to address this challenge. As illustrated in Fig. \ref{Device-Edge-Inference}, such techniques will partition the computation of DNN models, and offload part of the computation to the edge server. After being processed at the device, the amount of data to be offloaded can be reduced, and thus the offloading will be more efficient.

Neurosurgeon \cite{KanHau17} is a framework that can automatically partition DNN computation between the client and server at the granularity of the neural network layers. It adapts to various DNN architectures, hardware platforms, wireless networks, and server load levels, and intelligently partitions computation for best latency or best mobile energy. Both latency reduction and energy saving have been demonstrated. A similar framework, called Edgent, was proposed in \cite{LiZho18}. It partitions DNN computation between the device and edge, and further adopts an early-exit mechanism at a proper intermediate DNN layer to further reduce the computation latency.
In \cite{LiHu18}, a joint accuracy- and latency-aware (JALAD) execution framework was proposed to decouples a DNN among a client and a server. The trade-off between the model accuracy and execution latency was also investigated. How to apply these techniques to specific tasks of intelligent vehicles, such as perception, is a promising direction for further investigation.

\subsection{Vehicle as a Client}
\begin{figure}[t]
     \centering
     \begin{subfigure}{.45\textwidth}
         \centering
         \includegraphics[width=\textwidth]{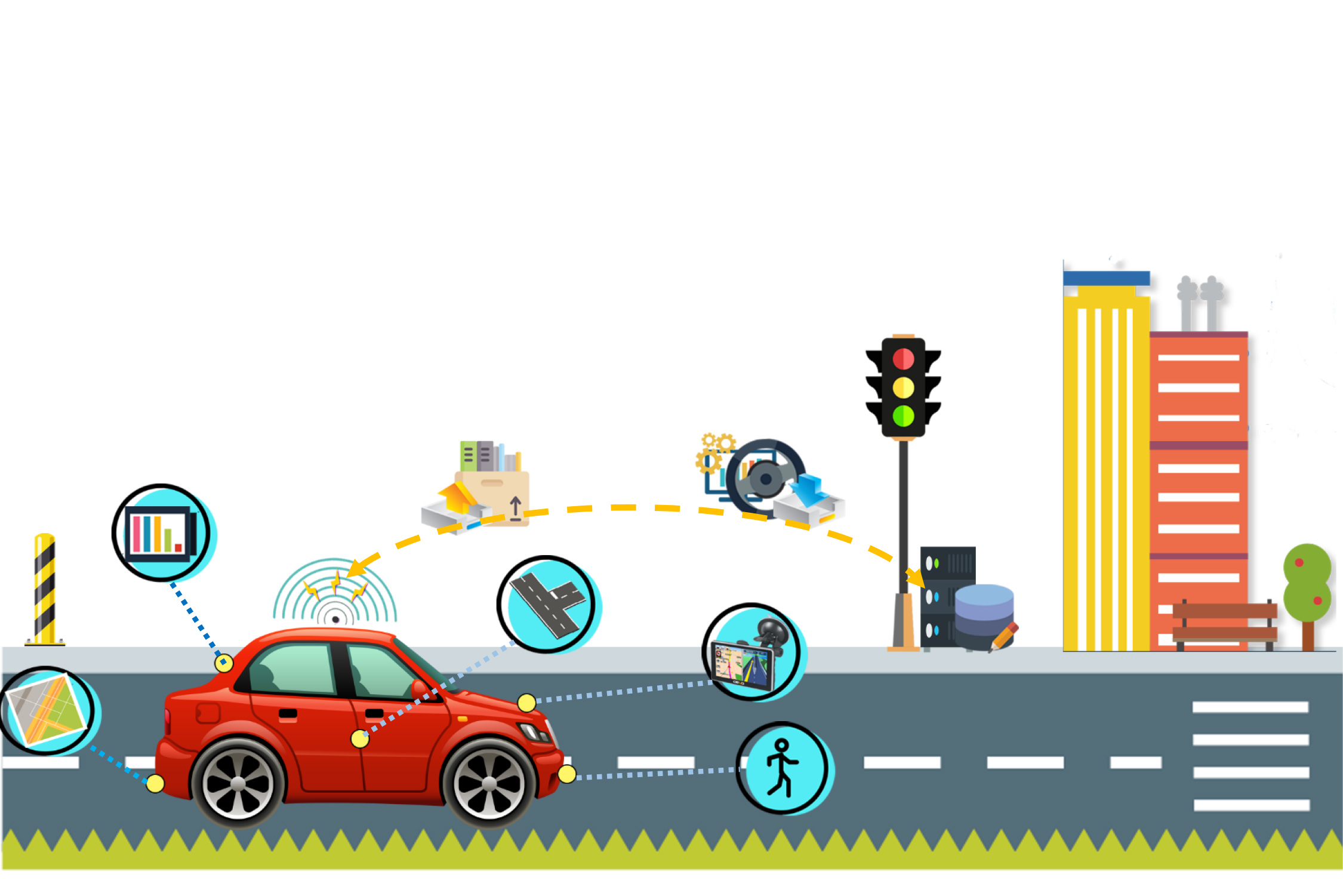}
         \caption{Perception}
         \label{fig:Perception}
     \end{subfigure}
     \hfill
     \begin{subfigure}{.45\textwidth}
         \centering
         \includegraphics[width=\textwidth]{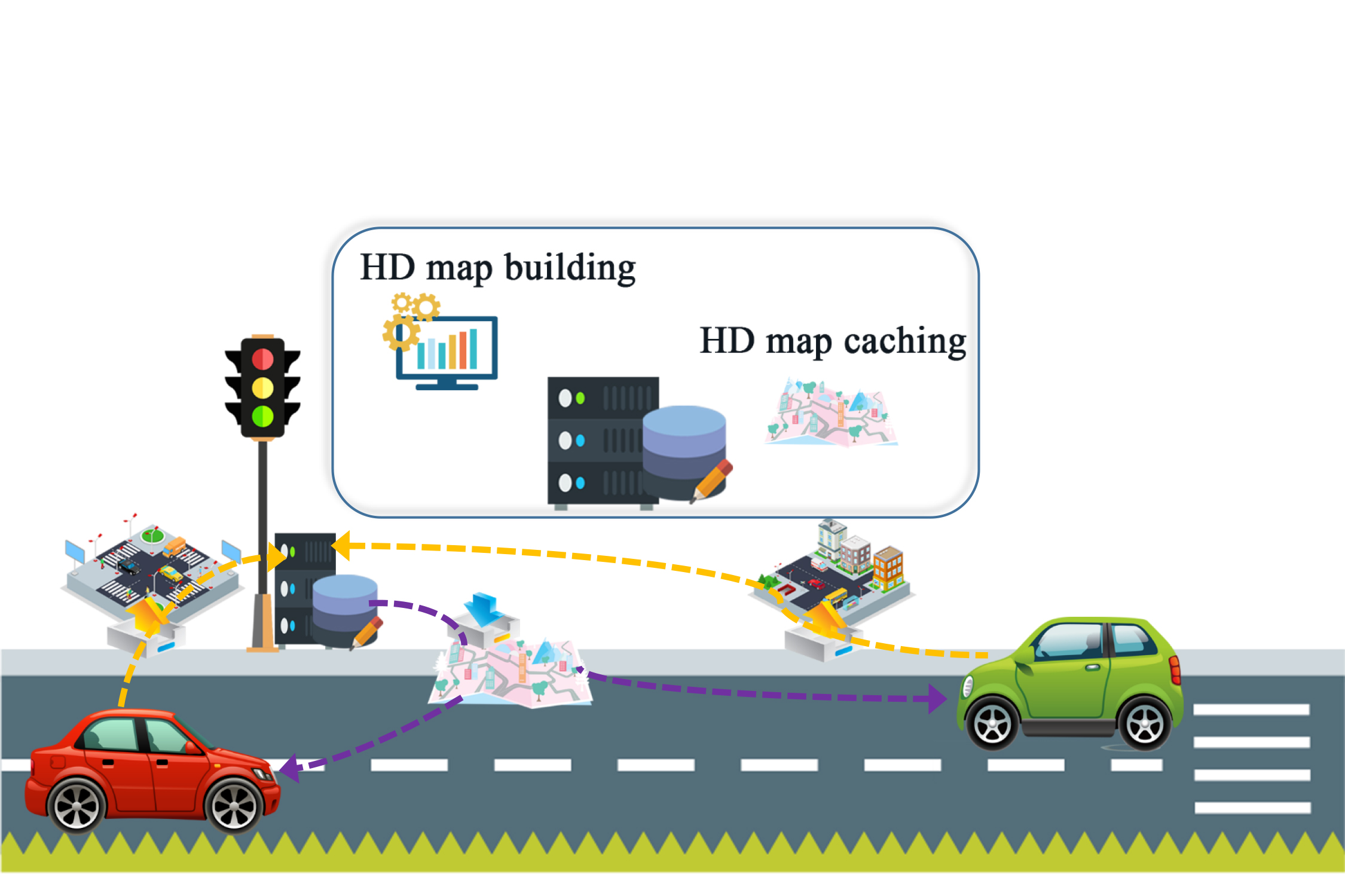}
         \caption{HD Mapping}
         \label{fig:HD_mapping}
     \end{subfigure}
     \hfill
     \begin{subfigure}{.45\textwidth}
         \centering
         \includegraphics[width=\textwidth]{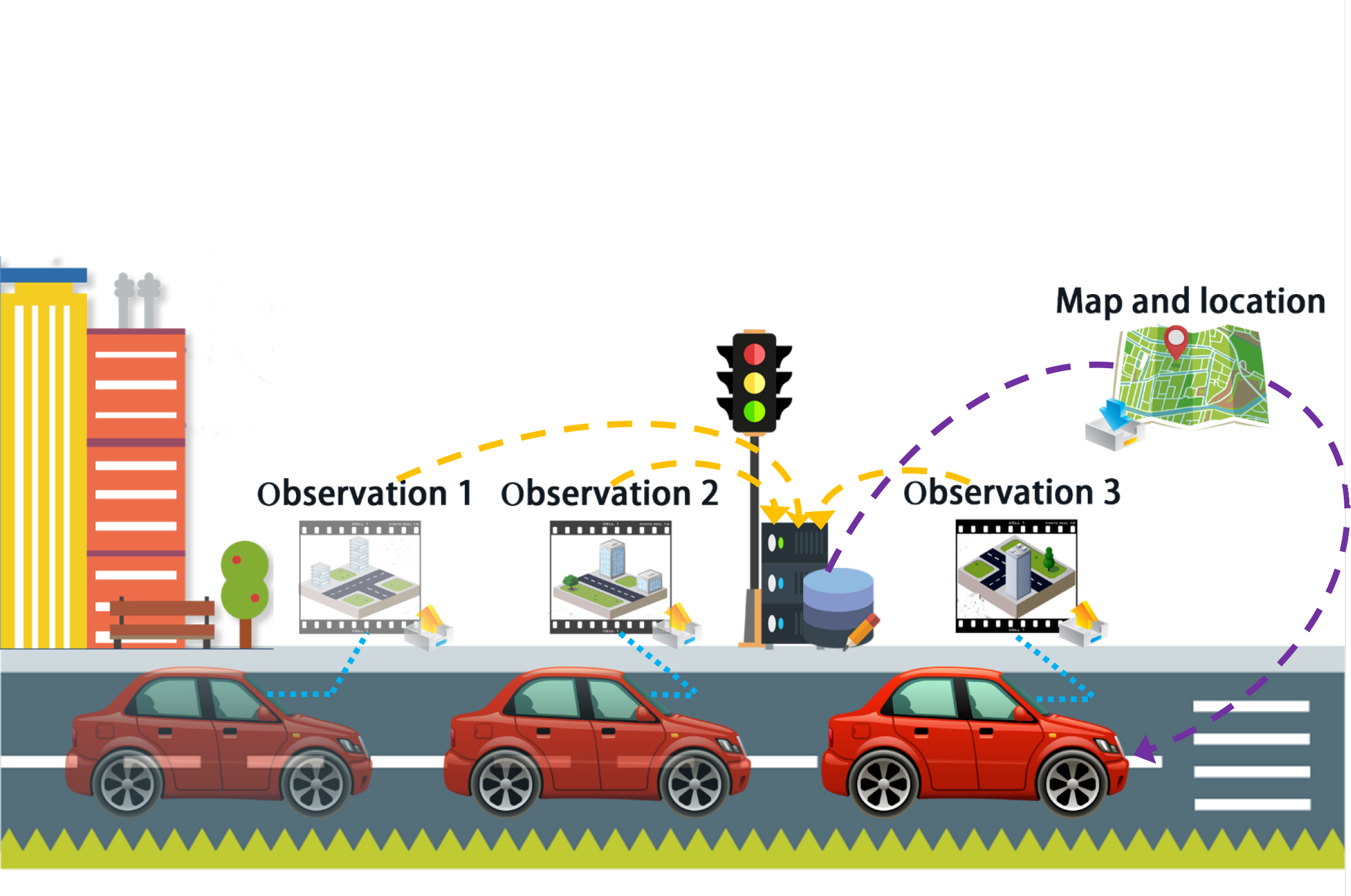}
         \caption{SLAM}
         \label{fig:SLAM}
     \end{subfigure}
        \caption{Three key tasks for intelligent vehicles.}
        \label{fig:Tasks}
\end{figure}

In this part, we present edge server assisted approaches for the three key tasks of intelligent vehicles, as illustrated in Fig. \ref{fig:Tasks}.

\subsubsection{Edge-Assisted Perception}
Perception tasks, such as object detection and tracking, can be assisted by edge servers, especially when powerful deep learning models are used. Two different offloading models can be considered, namely, binary and partial offloading \cite{MaoYou17}. For binary offloading, an offloading controller will determine whether the task will be executed by the on-board unit of the vehicle or be offloaded to the edge server, depending on factors such as the channel state, workload at the server, required computation intensity, etc. For partial offloading, the computation task is partitioned and executed at both the device and the edge server, and thus joint device-edge inference frameworks, e.g., \cite{KanHau17,LiZho18,LiHu18}, can be applied. Task migration among edge servers may be applied for mobility management \cite{SunZho17,OuyZho18}.

When designing offloading strategies, features of the specific algorithm should be carefully taken into account. For example, for localization, vision-based approaches enjoy highly parallel data-processing stages, such as feature extraction, disparity map generation, optical flow, feature match, and Gaussian blur \cite{ScaFra11}. Thus, they are more amenable for joint client-server processing by offloading part of the tasks to the edge server, which may provide abundant GPU resources. In comparison, LiDAR-based localization heavily uses the iterative closest point algorithm, which is hard to parallelize \cite{BesNei92}, and thus may be difficult to offload.

\subsubsection{Edge-Assisted Map Update}
To account for the dynamics of the environment, HD maps have to be refreshed timely, for which we need to collect fresh HD mapping data and detect changes from the fresh data. To build or update an HD map, the data collected in one area will be used only for that particular area. Thus, to alleviate the storage and communication burden, data collected by vehicles should be aggregated at the nearby edge servers, assisted by V2I communications. After enough fresh data are available, road change detection and road events (e.g., road closure) detection can be performed at the edge servers \cite{Jiao18}, e.g., applying the DNN-based methods \cite{AlcSte18}. Once each server builds its own map, multi-session mapping can be employed to combine multiple maps in a common metrical coordinate system \cite{McDKae11}. The updated map will be cached at the edge servers, which then help to notify the changes and distribute the map to vehicles in the coverage area.

\subsubsection{Edge-Assisted SLAM}

SLAM is a key technology for autonomous driving \cite{BreAls17}. Compared with mobile robots which are normally in the indoor environments, SLAM for autonomous driving is more challenging. To overcome the limitation of onboard computation, there have been studies on Cloud-based SLAM, i.e., to offload part of the computation load to the Cloud server. For example, Cloud framework for Cooperative Tracking And Mapping (C$^2$TAM) \cite{RiaCiv14} is a distributed framework where the expensive map optimization and storage is performed on the Cloud, while a light camera tracking client runs on a local computer. It applies the Parallel Tracking and Mapping (PTAM) algorithm, which has two parallel threads. On one hand, a geometric map is computed by non-linear optimization over a set of selected keyframes. This background process is able to produce an accurate 3D map at a low frame rate. On the other hand, a foreground tracking process is able to estimate the camera location at the frame rate assuming a known map. While the framework was developed assuming the Cloud platform, the experiment in the paper used a desktop as the ``Cloud''. Hence, it essentially is an edge-based SLAM method. Thus, this study demonstrated the feasibility of edge-assisted SLAM. Other existing studies on cloud-based SLAM for mobile robots, including DAvinCi \cite{AruEnt10} and Rapyuta \cite{MohHun15}, also provide valuable lessons for developing edge-assisted SLAM for intelligent IoV.

Edge servers may also assist cooperation among multiple vehicles for SLAM, which is called as \emph{centralized SLAM} \cite{BreAls17}. In this case, an edge server acts as a controller to aggregate and fuse the data before sending the results back to vehicles, via V2I communications. Different approaches have been proposed. Vehicles can first build sub-maps by themselves, which are then fused at the nearby edge server \cite{TaoHua10}. For this method, the relative locations of vehicles must be known. In another proposal \cite{GilRei10}, a multi-robot visual SLAM method was proposed, where data from multiple robots, i.e., observations and visual descriptors, are sent to a central agent to build a map, using a Rao-Blackwellized particle filter to estimate both the map and the trajectories of the robots. More investigation and testing will be needed along this line of research to evaluate its effectiveness.



\subsection{Vehicle as a Server}

With effective V2V communications, intelligent vehicles can share their sensing information and perception outputs. This enables various cooperative driving techniques \cite{KiaAug12}, for which edge servers at BSs or RSUs may act as coordinators.

\subsubsection{Cooperative Perception}
As discussed in Section \ref{EIS}, each intelligent vehicle is equipped with a combination of different sensors and cameras, but the sensing capability is fundamentally constrained by the inherent characteristic of each sensor, as well as the budget limit. For example, cameras have fundamentally limited visibility due to occlusions,
the sensing range, and extreme weather and lighting conditions. With IoV, vehicles are connected with each other, which provides the opportunity to cooperate for better sensing capabilities. Similar ideas have been applied for sensing and reusing available spectrum for cognitive radio in wireless communications systems \cite{ZhaMal09,ZhaLet09,LunKoi09}. Cooperative perception among multiple vehicles \cite{KimLiu15} improves situation awareness and perception capability, and provides traffic information beyond line-of-sight and field-of-view. It can be useful for situations such as hidden obstacle avoidance, safe lane-changing/overtaking, smooth braking/acceleration, etc.

Cooperative perception has received lots of attention, first for mobile robots for environmental surveillance \cite{MerCab06}, and recently for intelligent vehicles, e.g., cooperative localization \cite{LiNas13} and cooperative mapping \cite{LiTsu14}. In \cite{KimQin15}, a cooperative driving system based on cooperative perception was tested. A multimodal cooperative perception method was firstly developed, which provides a far-sight see-through, lifted-seat, satellite or all-around view to a driver. Cooperative driving by a see-through forward collision warning, overtaking/lane-changing assistance, and automated hidden obstacle avoidance was then tested through real-world experiments using four vehicles on the road. Recently, in \cite{QiuAhm17}, Augmented Vehicular Reality (AVR) was proposed to broaden the vehicle's visual range by sharing instantaneous 3D views of the surroundings with other nearby vehicles, assisted by effective V2V communications. There is no doubt that cooperative perception will play a key role in autonomous driving, and there are lots of research opportunities, e.g., developing communication-efficient approaches for information exchange between vehicles, leveraging recent advancements in distributed learning and estimation methods, etc.

\subsubsection{Crowd-Sourced Mapping}
While HD mapping is playing a significant role in autonomous driving, it is tedious and costly to construct an HD map. While autonomous driving companies could rely on their autonomous vehicles being tested on road to collect fresh data, the coverage is still limited. It is more efficient to work with automakers to get fresh map data from intelligent vehicles equipped with various sensors in a crowd-sourced manner. For example, with a large number of vehicles equipped with necessary hardware components for self-driving capability, it took Tesla only around four hours to collect 1 million miles of data \cite{Gia19}. In comparison, after years' driving tests, Waymo's self-driving fleet accumulated 10 million miles of data by October 2018 \cite{Kor18}, which Tesla could acquire within two days. While this is a rough comparison, the message is clear. The cache and computing resources at intelligent vehicles will be the basis for crowd-sourced map construction, while edge servers can act as local aggregators.

The constructed features in an HD map are managed by the layer-based map management system. This approach enjoys certain advantages. For example, the map editing such as addition, deletion, or correction, becomes efficient, and communication overhead can be reduced by downloading only the required layer. To deal with the increasing functional requirements of intelligent vehicles, new feature layers need to be added continuously. In \cite{KimCho18}, a crowd-sourced mapping process was proposed for constructing new feature layers for the HD map. First, data are acquired from multiple intelligent vehicles. Each vehicle then uses the acquired fresh data to build a new feature layer with the GraphSLAM approach. Next, these new feature layers are conveyed to a map cloud through a mobile network system, where they are integrated into a new feature layer in a map cloud. The feature integration can be performed at the local edge servers, to save communication overhead on the backbone network and reduce latency. While each vehicle may suffer from imperfections of the on-board sensors, combining many new feature layers acquired from many intelligent vehicles improves the accuracy.

\subsubsection{Multi-vehicle SLAM}

Cooperation between vehicles can be exploited to address the computational challenges in SLAM. Besides edge-assisted centralized SLAM mentioned in the previous subsection, \emph{decentralized SLAM} has also been proposed, where each vehicle builds its own decentralized map while communicating with the other vehicles \cite{SaeTre16}. In this way, vehicles can quickly update maps in case of sudden changes or to anticipate dynamic conditions, and can be robust to error/failure of any one of the vehicles. Decentralized SLAM is more challenging than centralized multi-vehicle SLAM, with difficulties including estimating relative poses of robots, uncertainty of the relative poses, updating maps and poses, complexity and communications issues, etc. Existing studies on decentralized SLAM are mainly for mobile robotics.

For decentralized SLAM, each vehicle has to integrate multiple local maps provided by the other vehicles to generate a global map. This is a challenging task as the required alignments or transformation matrices to relate different maps are in general unknown. One key principle is to identify the relative pose between spatial information from different vehicles, which can be handled by map merging, or map fusion. Map merging naturally involves two steps. First, it will find the required alignment between the local maps. It then merges maps by integrating information from the aligned maps into one map. Different approaches have been proposed, e.g., \cite{DinGiv12,BlaGon13,SaePau14}. To achieve real-time decentralized SLAM, communications among vehicles should be carefully considered. Depending on available bandwidth, different content can be exchanged, e.g., local maps \cite{WilDis02}, graph-based representations \cite{PfiSla07}, or topological maps \cite{VidBer11}. Out-of-sequence measurements, e.g., due to latency, should also be carefully handled \cite{BarChe04,HolSin12}. Compared with single-vehicle SLAM, more research and testing efforts will be needed for multi-vehicle SLAM.


\section{Conclusions}
\label{conclusion}

This paper introduced an edge information system for intelligent IoV, including edge caching, edge computing, and edge AI. Platforms, design methodologies, and key use cases were presented. Given the latest advancements and expected innovations in wireless communications, EIS is expected to become a key component of the information infrastructure that is needed to support intelligent IoV. Integrated with the evolving onboard processors, as well as Cloud computing platforms, it will support the acquisition, storing, and processing of the big data generated by intelligent vehicles.

There is no doubt that intelligent vehicles, especially autonomous vehicles, will revolutionize multiple economic sectors, and fundamentally change our daily life. During the last decade, we have witnessed a similar ground-breaking development, i.e., the rising of smartphones, which is supported by the mobile Internet and has affect every aspect of the society. We expect more dramatic and potentially revolutionary applications to occur in the following decades, with the development of intelligent vehicles and IoV. In particular, we envision that EIS will play a key role in this process. There are therefore abundant opportunities but also significant challenges. Efforts are hence needed from different players, including researchers, entrepreneurs, governments, policy makers, standardization bodies, etc., to help create the technologies and policies to meet the challenges ahead.

\bibliographystyle{IEEEtran}
\bibliography{Reference}

\end{document}